\documentclass[epj]{svjour}

\usepackage[intlimits]{amsmath}
\usepackage{amssymb}
\usepackage{multirow}
\usepackage{setspace}
\usepackage{graphicx}
\usepackage{textcomp}         
\usepackage{array}
\usepackage{slashed}
\usepackage[usenames,table]{xcolor}
\usepackage[colorlinks=true,linkcolor=blue,citecolor=blue,urlcolor=blue]{hyperref}
\usepackage[numbers,sort&compress]{natbib}

\begin{document}

\title{Hadronic bound-states in SU(2) from Dyson--Schwinger Equations}

\author{Milan Vujinovic${}^1$\thanks{\email{milan.vujinovic@uni-graz.at}}
   \and Richard Williams${}^2$\thanks{\email{richard.williams@theo.physik.uni-giessen.de}}}

\institute{Institut f\"ur Physik, Karl-Franzens--Universit\"at Graz, Universit\"atsplatz 5, 8010 Graz, Austria.
      \and Institut f\"ur Theoretische Physik, Justus-Liebig--Universit\"at Giessen, 35392 Giessen, Germany.}
\date{Received: date / Revised version: date}

\abstract{
By using the Dyson-Schwinger/Bethe-Salpeter formalism in Euclidean spacetime, we 
calculate the ground state spectrum of $J\leq 1$ hadrons in an SU(2) gauge 
theory with 2 fundamental fermions. We show that the rainbow-ladder truncation, 
commonly employed in QCD studies, is unsuitable for a description of an SU(2) 
theory. This we remedy by truncating at the level of the quark-gluon 
vertex Dyson-Schwinger equation in a diagrammatic expansion. Results 
obtained within this novel approach show good agreement with lattice 
studies. These findings emphasize the need to use techniques more sophisticated
than rainbow-ladder when investigating generic strongly interacting gauge theories. 
}


\maketitle

\section{Introduction}\label{sec:introduction}
Quantum Chromodynamics (QCD) is a strongly interacting gauge theory whose study 
has proven to be one of the most formidable challenges of modern theoretical 
physics. Whilst the high-energy regime of QCD is by now relatively well explored 
in terms of perturbation theory, the arguably more interesting (and 
intrinsically non-perturbative) phenomena such as dynamical chiral symmetry 
breaking and confinement are yet to be fully understood. 

One of the strategies which might lead to our better understanding of QCD is to 
investigate theories which are QCD-like, but have certain properties that make 
them technically less challenging than QCD itself. A prime example is provided 
by studies of SU(2) gauge theories with an even number of fermion flavors. 
Lattice simulations of these theories at non-zero chemical potential do not 
suffer from the sign problem, and such models thus provide ideal conditions to 
study the phase diagram of strongly interacting 
matter~\cite{Hands:1999md,Aloisio:2000rb,Aloisio:2000if,Hands:2000ei,Kogut:2001na,Hands:2001ee,Kogut:2002cm,Muroya:2002ry,Alles:2006ea,Hands:2006ve,Cotter:2012mb}. 

Here we wish to concentrate on the situation with two fundamentally 
charged Dirac 
fermions~\cite{Hands:1999md,Muroya:2002ry,Hands:2006ve,Cotter:2012mb}. Such a
theory may also be interesting in the context of a unified description of cold 
asymmetric Dark Matter (DM) and dynamical electroweak (EW) symmetry breaking
~\cite{Lewis:2011zb,Hietanen:2014xca,Hietanen:2013fya}, wherein the ground state 
hadronic spectrum at $T=0$, $\mu = 0$ is of great importance. It is exactly this 
hadronic spectrum that will be the central focus of our study.

In this paper we use the non-perturbative, continuous and covariant 
formalism of Dyson-Schwinger (DSE) and Bethe-Salpeter (BSE) equations in 
Euclidean spacetime
~\cite{Roberts:1994dr,Alkofer:2000wg,Fischer:2006ub,Cloet:2013jya}. When applied 
to QCD, the most common truncation one can make is that of 
rainbow-ladder (RL), wherein the quark-antiquark interaction kernel is 
replaced by a dressed one gluon exchange. It is the simplest 
approximation scheme that respects the axial-vector Ward-Takahashi identity 
(axWTI), thus preserving the chiral properties of the theory and the
(pseudo)-Goldstone boson nature of light pseudoscalar mesons. With a judicious 
choice of model dressing functions, the RL truncation has been applied 
relatively successfully to QCD phenomenology for both mesons~\cite{Maris:1997hd,Maris:1997tm,Maris:1999nt,Krassnigg:2009gd,Krassnigg:2009zh,Blank:2010pa,Krassnigg:2010mh,Blank:2011ha,Roberts:2011cf,Fischer:2014xha,Hilger:2014nma,Fischer:2014cfa} 
and baryons~\cite{Eichmann:2009qa,Nicmorus:2010sd,SanchisAlepuz:2011aa,Eichmann:2011vu,Sanchis-Alepuz:2013iia}. 

However, as we will show in this paper, the RL truncation performs unsatisfactorily 
when adapted to an SU(2) theory with 2 fundamental flavors, even though the 
theory is expected to have QCD-like dynamics. We discuss possible reasons for
this in more detail in Section \ref{sec:framework}. Here we only comment that we 
strongly believe that (most) of the inadequacy of RL method comes from its weak 
connection to the underlying gauge sector. Remedying this requires the use of 
beyond rainbow ladder (BRL) techniques, with our preference towards those based on 
the diagrammatic expansion of quark-gluon vertex 
DSE~\cite{Munczek:1994zz,Bender:1996bb,Bender:2002as,Watson:2004kd,Watson:2004jq,Bhagwat:2004hn,Matevosyan:2006bk,Fischer:2009jm,Windisch:2012de,Gomez-Rocha:2014vsa,Williams:2014iea}.
Whilst there are other BRL methods available~\cite{Fischer:2005en,Fischer:2007ze,Fischer:2008sp,Fischer:2008wy,Chang:2009zb,Chang:2011ei,Heupel:2014ina}, 
we choose the diagrammatic approach as it makes it easier to study the influence 
of the gauge sector on hadronic observables. Our aim in this paper is thus not 
only to provide a continuum calculation complimentary to the lattice 
investigations of~\cite{Lewis:2011zb,Hietanen:2014xca}, but also to explicitly 
demonstrate the importance of using BRL methods when studying generic strongly 
interacting theories. 

This manuscript is organised as follows. In Section~\ref{sec:framework} we 
discuss the DSEs relevant for our calculation, and also describe in some detail 
the approximations and model inputs we employ. In 
Section~\ref{sec:spacelikeboundstates} we describe the extrapolation procedures 
used to obtain hadron masses, and provide estimates for errors coming from 
extrapolation. The results are disscused and compared to relevant lattice data 
in Section~\ref{sec:results}. We conclude in Section~\ref{sec:conclusion}.  

\begin{figure}[!b]
\begin{center}
\includegraphics[height=2cm]{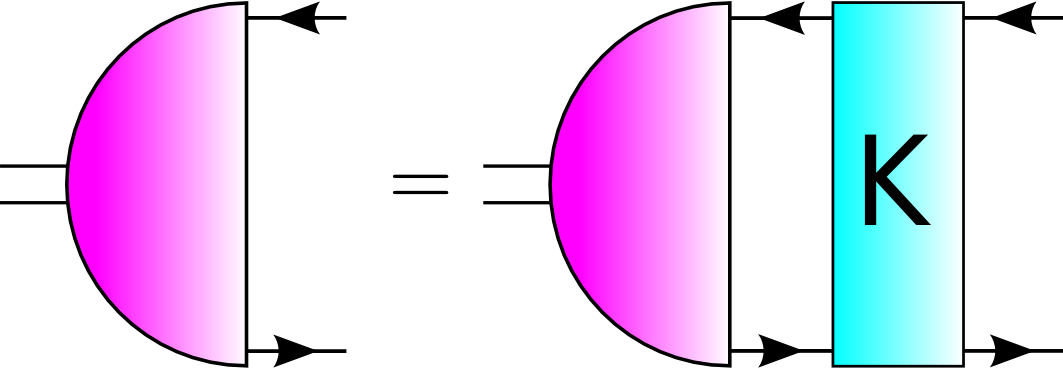}
\caption{The Bethe--Salepter equation for the meson.}\label{fig:mesonbse}
\end{center}
\end{figure}
\section{Framework}\label{sec:framework}
In a theory with 2 colors, both mesons and baryons (diquarks) can be described 
in terms of a two-body Bethe-Salpeter equation. For the meson
\begin{align}\label{eqn:mesonbse}
\left[\Gamma_M(p, P)\right]_{ij} = \int_k\left[K(p,k,P)\right]_{ik;lj} \left[\chi_M(k,P)\right]_{kl} \;,
\end{align}
where $\int_k$ stands for $\int d^4k/(2\pi)^4$ and $\Gamma_M(p,P)$ is the 
meson amplitude with appropriate $J^{PC}$ quantum numbers, relative 
momentum $p$ and total momentum $P$, and the meson wavefunction is $\chi_M(k,P)=S(k_+)\Gamma_M(k,P)S(k_-)$.
  The quark propagators are $S(k_\pm)$, at 
momenta $k_+ = k + \eta P$ and $k_- = k - (1-\eta)P$, with $k$ the loop momentum 
and $\eta\in[0,1]$ the momentum partition factor. In a covariant study, the 
results are independent of $\eta$: for concreteness, we work with 
$\eta = 1/2$. The final ingredient in Eq.~\eqref{eqn:mesonbse} is the 
quark-antiquark 4-point interaction kernel $K(p,k,P)$. A diagrammatic 
representation of Eq.~\eqref{eqn:mesonbse} for mesons is given in 
Fig.~\ref{fig:mesonbse}.

\begin{figure}[!t]
\begin{center}
\includegraphics[height=1.0cm]{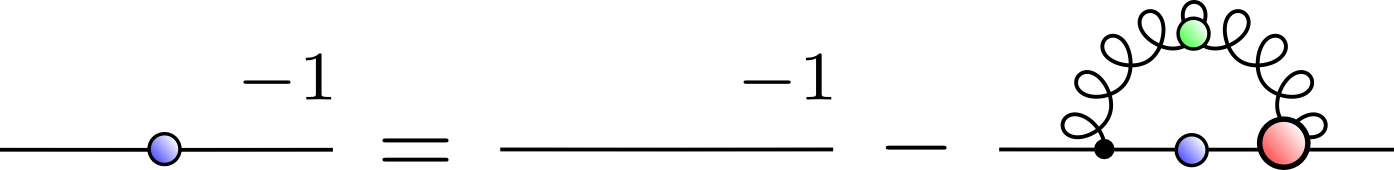}
\caption{The Dyson--Schwinger equation for the quark propagator. Straight lines
are quarks, wiggly ones gluons. Filled circles indicate dressed propagators and vertices.}
\label{fig:quarkdse}
\end{center}
\end{figure}

In order to solve the BSE, one clearly needs as input the 
quark propagator $S(p)$. This Green's function is decomposed as 
\begin{align}\label{eqn:quark}
S^{-1}(p) = Z_f^{-1}(p^2)\left[ i\slashed{p} + M(p^2)\right] \;,
\end{align}
with $Z_f(p^2)$ the quark wavefunction and $M(p^2)$ the dynamical quark mass. 
The tree-level form is given by $S^{-1}_0(p) = i\slashed{p} + Z_m m$, where
$Z_m$ is the quark mass renormalisation constant. The quark propagator satisfies 
its own DSE, see Fig.~\ref{fig:quarkdse}, and is given by
\begin{align}\label{eqn:quarkdse}
S^{-1}(p) & = Z_2S_0^{-1}(p) \\
          & + g^2 Z_{1f}C_F\int_k \gamma^\mu S(k+p)\Gamma^\nu(k+p,p)D_{\mu\nu}(k)\;.\nonumber
\end{align}
Here, $\Gamma^\nu(p,k)$ and $D_{\mu\nu}(k)$ are the full quark-gluon 
vertex and gluon propagator, respectively. Renormalisation constants of the
quark field and quark-gluon vertex are $Z_2$ and $Z_{1f}$. They are 
related through a Slavnov-Taylor identity which takes a simple form when 
employing a miniMOM scheme~\cite{vonSmekal:2009ae} in Landau gauge, 
$Z_{1f} = Z_2/\widetilde{Z}_3$ with $\widetilde{Z}_3$ the renormalisation of the 
ghost propagator.

The $4$-point interaction kernel $K(p,k,P)$ of 
Eq.~\eqref{eqn:mesonbse} is connected to the self-energy part $\Sigma(p)$ of 
quark propagator DSE through the axial-vector Ward-Takahashi identity (axWTI)
\begin{align}\label{eqn:axWTI}
[\Sigma(p_+)\gamma_5 & + \gamma_5\Sigma(p_-)]_{ij} = \\
                    &\int_k \left[K(p,k,P)\right]_{ik;lj} \left[\Sigma(k_+)\gamma_5  + \gamma_5\Sigma(k_-)\right]_{kl}\;.\nonumber
\end{align}
This identity encodes the chiral properties of the theory, and severely 
constrains the form of the BSE interaction kernel once an approximation for the 
quark DSE has been chosen. A direct connection is provided through the action of
`cutting' internal quark lines~\cite{Munczek:1994zz,Bender:1996bb}. 

\begin{figure}[!b]
\begin{center}
\includegraphics[height=2.0cm]{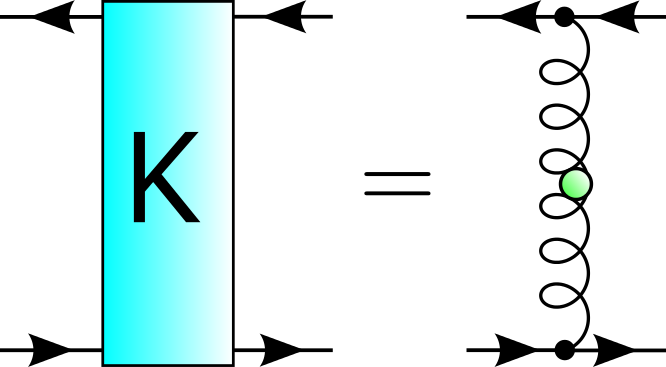}
\caption{The truncated two-body kernel in rainbow-ladder approximation.}
\label{fig:kernelRL}
\end{center}
\end{figure}
\subsection{Rainbow-ladder}\label{sec:rl}
The `rainbow' part of RL truncation refers to the replacement of the full 
quark-gluon vertex in Eq.~\eqref{eqn:quarkdse} by
\begin{align}\label{eqn:rainbow}
\Gamma^\nu(k+p,p)\rightarrow \lambda(k^2)\gamma^\nu\;,
\end{align}
i.e. its tree-level form augmented by a model dressing function, $\lambda(k^2)$, 
that is a function of the gluon momentum alone. The corresponding 
axWTI-preserving approximation for BSE kernel is that of one gluon exchange 
(the `ladder'), which we show diagrammatically in Fig.~\ref{fig:kernelRL}. 

In the RL approach, the model dressing function $\lambda(k^2)$ of 
Eq.~\eqref{eqn:rainbow} is often combined with the dressing of the gluon 
propagator $D_{\mu\nu}(k^2)$ into a single model function, constructed to 
reproduce correctly some hadronic observables, usually $m_\pi$ and $f_\pi$. 
Whilst this method has shown considerable success in QCD phenomenology
(see e.g.~\cite{Eichmann:2008ae, Qin:2011xq} for some of the limitations of the 
model), in an SU(2) theory the approach seems rather unsuitable, especially in
the $1^{++}$ channel: see Table~\ref{tab:comparewithlattice} for details.

There are two primary reasons why RL will not perform satisfactorily in a 
generic strongly-interacting theory. One reason is with regards to its very 
limited interaction structure ($\gamma^\nu \times \gamma^\mu$) which offers no 
variation in interaction strength across different meson channels. The second 
is that the connection to the underlying gauge dynamics is typically lost 
in the construction of an effective quark-gluon interaction; this prevents
adequate rescaling of parameters such as $g^2N_c$ that cannot be translated
into a re-parameterization of an effective model. 

\subsection{Beyond rainbow-ladder}\label{sec:beyondrl}
\begin{figure}[!b]
\begin{center}
\includegraphics[scale=0.6]{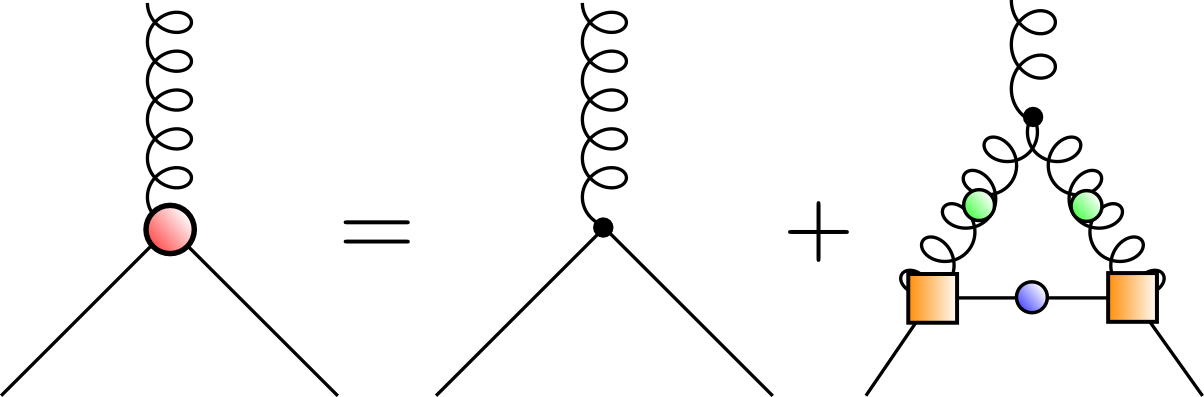}
\caption{The truncated DSE for the quark-gluon vertex. The orange square denotes
the internal QG vertex model, according to Eq.~\eqref{eqn:rainbow}.}\label{fig:qgdse}
\end{center}
\end{figure}

\begin{figure}[!h]
\begin{center}
\includegraphics[scale=0.6]{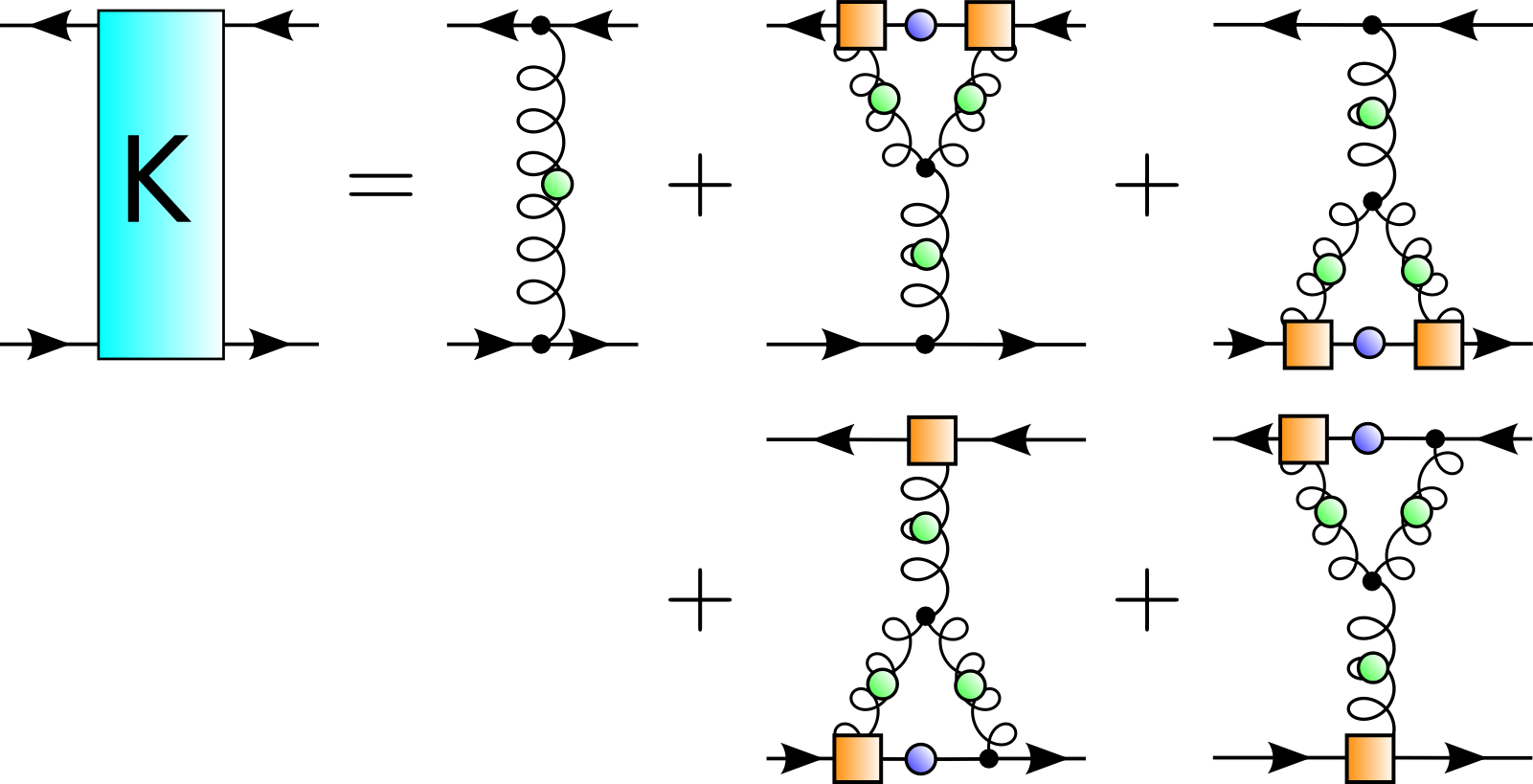}
\caption{The truncated two-body kernel beyond rainbow-ladder approximation.}\label{fig:kernel2body}
\end{center}
\end{figure}

A BRL approach which is well suited for studying the influence of underlying 
Yang-Mills sector on the hadronic observables is based on the quark-gluon 
vertex~\cite{Skullerud:2002ge,Skullerud:2003qu,Kizilersu:2006et,Bhagwat:2004hn,
Windisch:2012de,Rojas:2013tza,Williams:2014iea,Aguilar:2014lha,Mitter:2014wpa}. Here, we
focus on the truncated form of the DSE~\cite{Williams:2014iea} shown in
Fig.~\ref{fig:qgdse}. Within this approximation, only the so-called non-Abelian 
(NA) diagram is kept in the quark-gluon vertex self-energy. The truncated kernel, consistent with 
constraints from chiral symmetry, is shown in Fig.~\ref{fig:kernel2body}.      

So that the Bethe-Salpeter equation can be tackled, the evaluation of a  
fully self-consistent quark-gluon vertex is not performed. That is, the 
full calculated vertex (denoted by a red filled circle in Fig.~\ref{fig:qgdse})
is not back-coupled into the non-Abelian diagram. Instead, the internal 
vertices (orange squares in Fig.~\ref{fig:qgdse}) are modelled by the Eq.~\eqref{eqn:rainbow}
with $\lambda(k^2)$ constructed such that it strongly 
resembles the tree-level projection of the full quark-gluon vertex at each 
iteration step; essentially, it depends upon a function $\Lambda(M_0)$ that encodes the 
interaction strength in terms of the dynamically generated quark mass. We used the 
parametrisation Eq.~(21) of Ref.~\cite{Williams:2014iea}, with
modifications that account for the change $N_c=2$ and the rescaling of 
the gauge coupling $g^2$ ($g^2 N_c$ is left invariant). For $\Lambda(M_0)$  we 
use the functional form given in Eq.~(22) of~\cite{Williams:2014iea}
with parameters $a \simeq 2.44, b \simeq 1.79, c\simeq -0.20, d \simeq 0.30$.  The
procedure described therein is used for solving the resultant coupled DSE system of a quark 
propagator and quark-gluon vertex.

\begin{figure}[!b]
\begin{center}
\includegraphics[width=0.93\columnwidth]{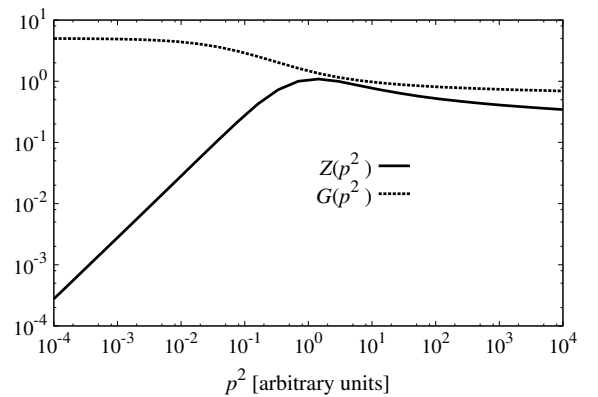}
\caption{Ghost ($G$) and gluon ($Z$) dressing functions employed in our calculations. 
The momentum $p^2$ is in arbitrary units: scale setting procedure is described in Section~\ref{sec:analysis}.}\label{fig:ghostandgluon}
\end{center}
\end{figure} 

We emphasize here that this model is, in a sense, highly constrained. Namely, once 
the input for the ghost and gluon propagators (which we will discuss
shortly) and the truncation of the quark-gluon vertex DSE have been chosen, all other
parts of the calculation are fixed. The BSE kernel follows
from the axWTI, and the model dressing $\lambda(k^2)$ of Eq.~\eqref{eqn:rainbow}
follows from the tree-level projection of the full quark-gluon vertex. We will
re-iterate this point in Section \ref{sec:modeldependence}, when we provide
an estimation of the model dependence.

The final ingredient which we need to specify in our calculation is the gluon propagator 
$D_{\mu\nu}(k)$. We work in Landau gauge, where this Green's function takes the 
form 
\begin{align}\label{eqn:gluon}
D_{\mu\nu}(k) = T_{\mu\nu}(k)\frac{Z(k^2)}{k^2}\;,
\end{align}   
with $T_{\mu\nu}(k) = \delta_{\mu\nu} - k_\mu k_\nu/k^2$ the transverse 
projector with respect to momentum $k$. The gluon dressing function which we use 
is plotted in Fig.~\ref{fig:ghostandgluon}. The details of this function and its
parametrisation can be found in~\cite{Eichmann:2014xya}. We point out that the 
gluon which we employ corresponds to a quenched DSE calculation. Ignoring the 
back-reaction of quarks onto the Yang-Mills sector is usually considered a good 
approximation for theories with QCD-like dynamics, as the corresponding effect 
on `observables' like the chiral condensate, $f_\pi$ and others is quite 
small~\cite{Fischer:2003rp}. However, the quenched approximation should be 
reconsidered in theories which have (nearly) conformal, or `walking' dynamics. 
Walking dynamics arises naturally in models with a relatively large number of
light fundamentally charged fermions \cite{Caswell:1974gg,Appelquist:1986an,Appelquist:1987fc,Appelquist:2007hu,Bursa:2010xn,Hasenfratz:2011xn,Cheng:2011ic,Lin:2012iw,Hopfer:2014zna},
or fermions belonging to higher-dimensional representations of the gauge group \cite{Sannino:2004qp,Dietrich:2005jn,Dietrich:2006cm,Catterall:2007yx,Maas:2011jf,DeGrand:2011qd,Bursa:2011ru,Fodor:2012ty,DeGrand:2013uha}.    

\section{Bound-states from space-like $P^2$}\label{sec:spacelikeboundstates}
One of the consequences of working with Euclidean spacetime is that access to 
time-like quantities, such as masses of bound-states, requires an analytic 
continuation of the component Green's functions to complex momenta.
Whilst this is only a minor technicality thanks to many well-established 
techniques in the literature
~\cite{Fischer:2008sp, GimenoSegovia:2008sx, Krassnigg:2009gd, Strauss:2012dg, Windisch:2013dxa}, 
there are situations in which existing methods do not apply, or which are simply 
too complicated to implement. In this case, indirect methods can be employed 
that enable access to a limited number of time-like 
quantities~\cite{Bhagwat:2007rj, Dorkin:2013rsa}.

In the next two sections we describe two techniques that have been widely used, 
and compare their performance in cases where direct analytic continuation is 
possible. This provides an estimate of the methods applicability to the study at 
hand.

\subsection{Eigenvalue extrapolation}
There are several means by which the mass spectrum of the BSE can be obtained. 
The most often used is through solution of Eq.~\eqref{eqn:mesonbse}, written as
a matrix equation for simplicity
\begin{align}
\Gamma_i = \lambda\left(P^2\right) K_{ij}\Gamma_j\;.
\end{align}
This has solutions at discrete values of the bound-state's total momentum 
$P^2=-M_i^2$. By introducing the function $\lambda(P^2)$ on the right, we obtain 
an eigenvalue equation whose bound-state solution correspond to 
$\lambda\left(P^2\right)=1$.

Since $\lambda(P^2)$ is a continuous function of 
$P^2$, one can conceive that its continuation from spacelike $P^2>0$ to timelike 
$P^2<0$ may be obtained through appropriate function fitting and extrapolation.
The transformation of the eigenvalue $g\left(\lambda\right) =
1 - 1/\lambda$, see Ref.~\cite{Blank:2011ha}, removes a considerable degree of intrinsic curvature
in the region close to the pole, rendering simple linear extrapolation viable provided the extrapolation is not \emph{far}.

In the top panel of Fig.~\ref{fig:extrapolationRL} we show the eigenvalue 
extrapolation of $\lambda\left(P^2\right)$ for various $J^{PC}$ states. The data 
is first transformed via $g\left(\lambda\right)$, before a linear fit 
$f\left(P^2\right) = a + b x$ is performed. Finally, we plot the inverse function of $g$, 
$\lambda_{\mathrm{fit}}=g^{-1}\left(f\left(P^2\right)\right)$ as solid lines. 
Exact results, obtained via calculation in the complex plane, are included as 
labelled points.

\subsection{Inverse vertex function extrapolation}\label{sec:vertexfunction}
The second means to obtain the mass spectrum employs instead the inhomogeneous 
BSE for the vertex function $\Gamma_i$
\begin{align}\label{eqn:vertexfunction}
V_i = V^{(0)}_i + K_{ij}V_j\;.
\end{align}
The obvious difference between this and the homogeneous BSE is the inhomogeneous 
term $\Gamma^{(0)}_i$. Its introduction leads to several important changes to 
the solution. Setting the relative momentum $p$ to zero, for convenience, we 
observe the appearance of poles 
\begin{align}\label{eqn:inhomogeneousatpole}
V_i(P^2) \sim \frac{1}{P^2+M^2}\;,
\end{align}
as one approaches the bound-state $P^2\sim-M^2$. Then, the determination of a 
bound-state mass is reduced to looking for zeros in $1/V_i$. Typically, the 
leading amplitude is used as point of reference, and one employs the method of 
bi-conjugate gradient (stabilised) for solution.   

\begin{figure}[!t]
\begin{center}
\includegraphics[width=0.93\columnwidth]{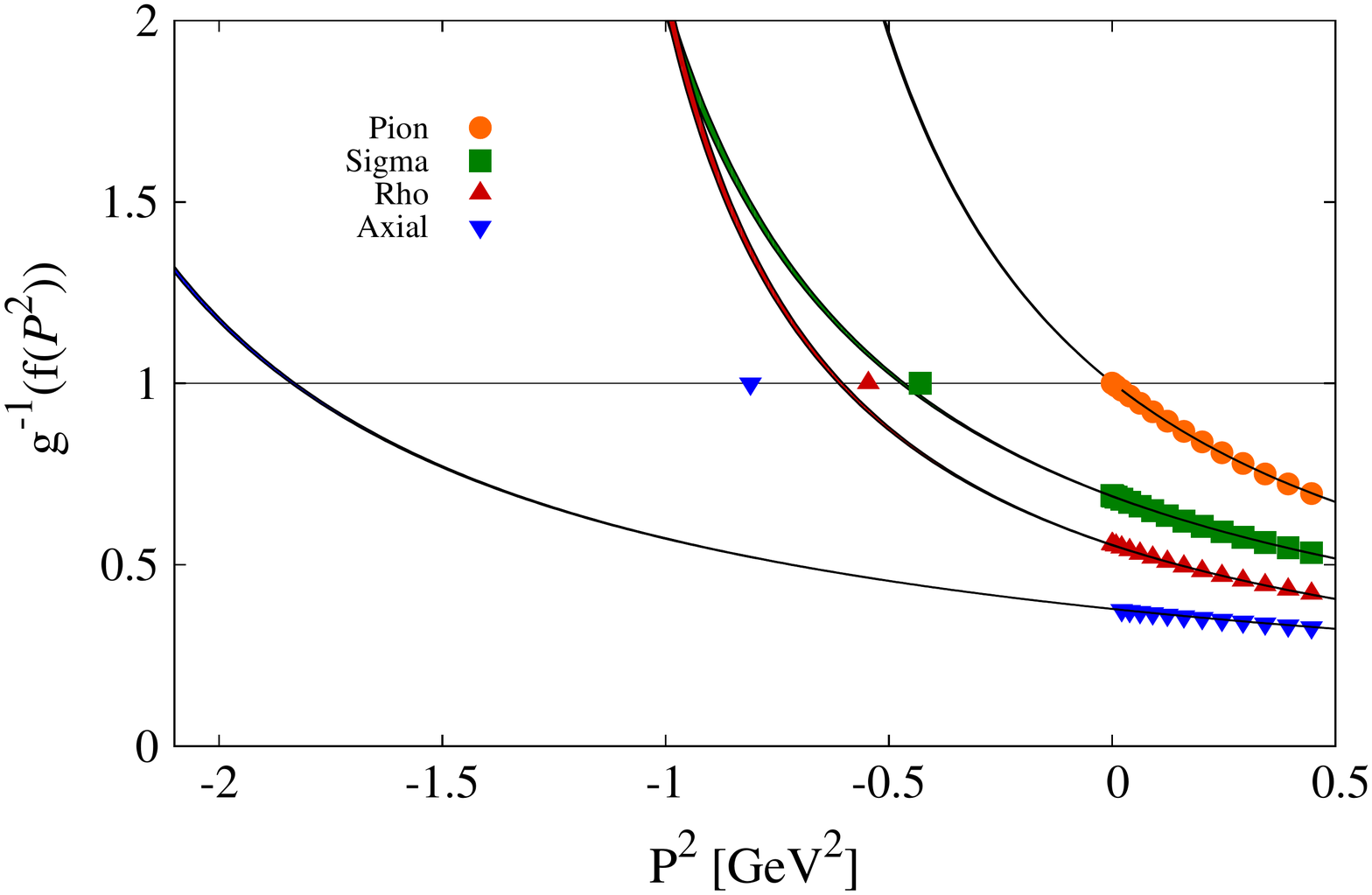}
\includegraphics[width=0.93\columnwidth]{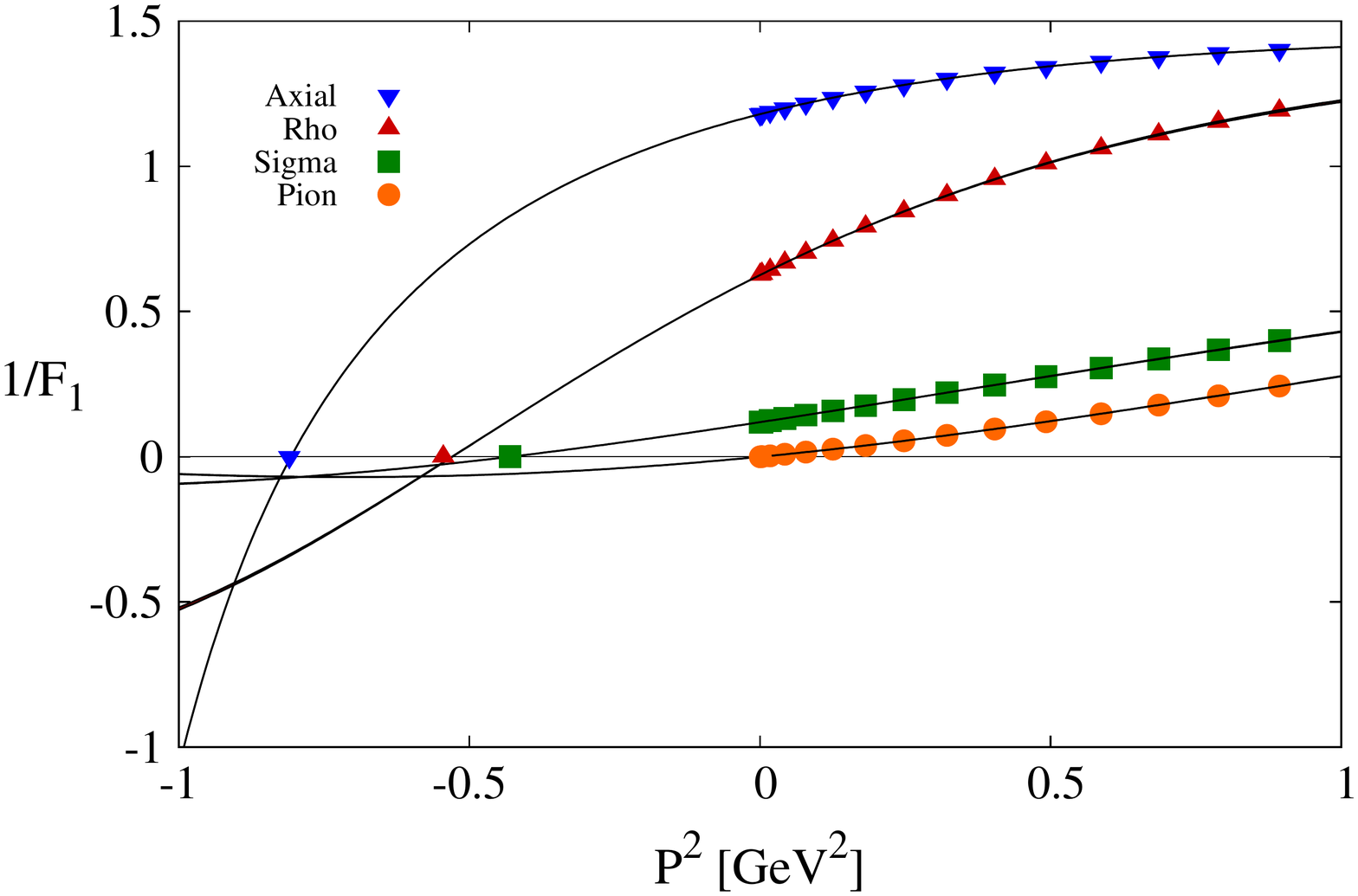}
\caption{Eigenvalue (\textit{top}) and vertex pole (\textit{bottom}) 
extrapolation from $P^2>0$ to the time-like region. The known result, 
obtained by direct analytic continuation, is given by labeled points for 
comparison.\label{fig:extrapolationRL}}
\end{center}
\end{figure}

Restricting ourselves to spacelike momenta $P^2$ requires once more the use of 
fit functions and extrapolation. Here, the most useful are rational polynomials
\begin{align}\label{pade}
R^{n,m}(x) = \frac{\sum_{i=0}^n a_i x^i}{1+\sum_{i=1,m}b_i x^i}\;.
\end{align}
Note, that since the coefficients $a_i$, $b_i$ are obtained through 
least-squares fitting, the resulting function is not a true Pad\'e. Regardless, 
the procedure appears quite reliable as can be seen in the bottom panel of 
Fig.~\ref{fig:extrapolationRL}.

\begin{table}[!h]
\begin{center}
\caption{Results for vertex pole extrapolation for QCD rainbow-ladder in the 
chiral limit, compared with the result computed through direct analytic 
continuation. All units are in MeV. The points $P^2$ are taken from the region $(0,L)$; 
the errors on extrapolated results come from the fitting procedure.}
\label{tab:resultsrl}
\begin{tabular}{c|c|cc}
\hline\hline\noalign{\smallskip} 
$J^{PC}$    &  calc      & $R^{(2,2)}$ ($L=0.5$) & $R^{(2,2)}$ ($L=1.0$) \\
\noalign{\smallskip}\hline
$0^{-+}$    &    0       &        $1$            &        $1$             \\      
$0^{++}$    &   658      &        $657(23)$      &        $656(23)$       \\
\noalign{\smallskip}\hline
$1^{--}$    &  738       &        $731(27)$      &        $728(27)$        \\
$1^{++}$    &  900       &        $899(33)$      &        $899(33)$           \\
\noalign{\smallskip}\hline\hline
\end{tabular}
\end{center}
\end{table}

We summarize our results for vertex pole approximation in 
Table.~\ref{tab:resultsrl}. The results obtained with eigenvalue 
extrapolation are not quoted as the method performs rather poorly, especially in
the $1^{++}$ channel (see top panel of Fig.~\ref{fig:extrapolationRL}). In either
of the extrapolation techniques there are two principal sources of uncertainty 
for the mass values. One comes from the fitting procedure, since 
the fit function coefficients ($a_i, b_i$ of Eq.~\eqref{pade} for vertex pole method)
come with their own error bars. These errors are straightforward to quantify, and the
resulting uncertainties for meson masses are quoted in parentheses in Table.~\ref{tab:resultsrl}.

A second source of errors has to do with the applicability of the extrapolation procedure, as
one would expect the whole method to become less reliable as one probes deeper into the $P^2 <0$ 
region (i. e. the technique is less reliable for heavier mesons). Although it is very hard to
quantify this, a comparison with exact results suggests that these effects are quite small for
the inverse vertex approximation. 
In light of other systematic errors, present in both the continuum and lattice investigations of 
the SU(2) gauge theory, we will ignore this uncertainty in Section \ref{sec:results}. 
As an additional check on the extrapolation method, we performed calculations with different fit
ranges for $P^2$, with the total momentum sampled in the region $(0,L)$ (in GeV$^2$), and with $L$ given
in the table. In the next section we employ the method with $L=0.5$, which appears empirically to have the
best performance.

\section{Results}\label{sec:results}
\subsection{Estimation of model dependence}\label{sec:modeldependence}
As already highlighted, the majority of model dependence stems from 
the truncation of the quark-gluon vertex DSE. Other parts of the calculation are constrained 
either by the underlying gauge dynamics (i.e. the ghost and gluon propagator which  
are taken from appropriate lattice or continuum calculations) or by chiral 
symmetry (in the process of truncating the BSE kernel). Thus, we can test the sensitivity
of the truncation by varying the solution of the quark-gluon vertex within the constraints
imposed by chiral symmetry breaking and the axWTI.

\begin{figure}[!t]
\begin{center}
\includegraphics[width=0.93\columnwidth]{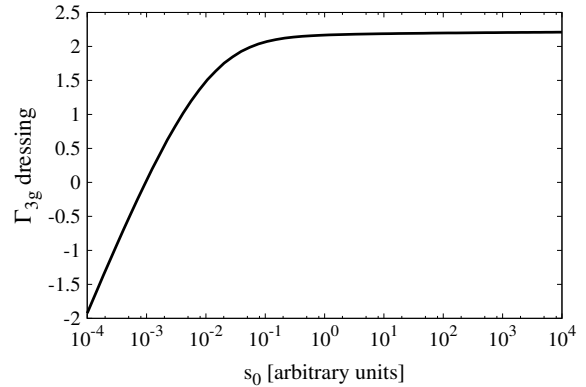}
\caption{Dressing for
the three-gluon vertex, with $s_0 = (1/6)\cdot(p_1^2 + p_2^2 + p_3^2)$ and $a=s=0$, see Eq. (50) of \cite{Eichmann:2014xya}. The momentum variable
$s_0$ is in arbitrary units: scale setting procedure is described in Section~\ref{sec:analysis}.}\label{fig:threegluon}
\end{center}
\end{figure} 

\begin{figure}[!b]
\begin{center}
\includegraphics[height=2cm]{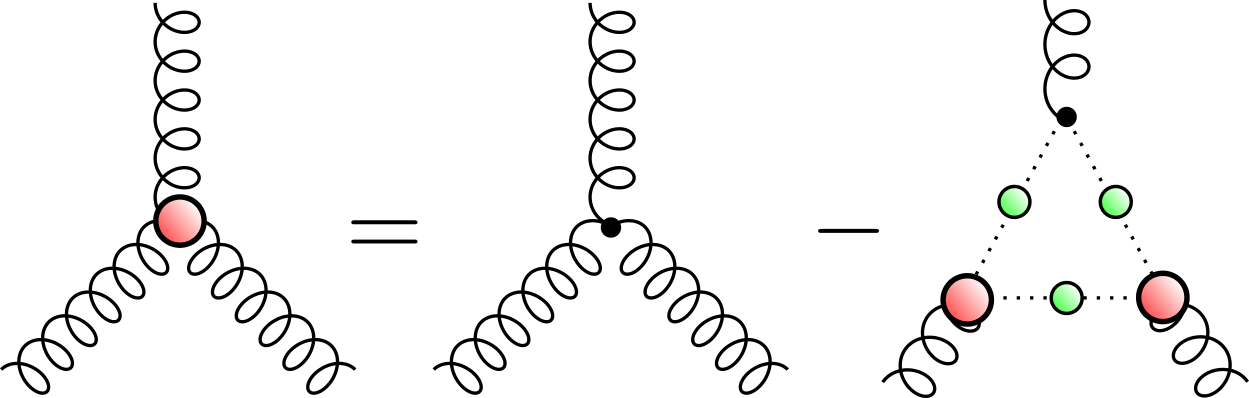}
\caption{The truncated DSE for the three-gluon vertex. To ensure that bose-symmetry is maintained the right-hand side is averaged over all cyclic permutations.}\label{fig:3gvertexghostonly}
\end{center}
\end{figure}

The natural step is to dress the three-gluon vertex. This is motivated 
by both the 3PI formalism~\cite{Berges:2004pu} and through the effective 
resummation of neglected diagrams in the full DSE for the quark-gluon vertex. 
This in turn enables us to give a rough estimate as to the impact of including
additional corrections on our results. It is sufficient to describe the full
three-gluon vertex in Landau gauge by its tree-structure and one function of a 
symmetric variable $s_0 = (1/6)\cdot(p_1^2 + p_2^2 + p_3^2)$~\cite{Eichmann:2014xya}
\begin{align}\label{eqn:3gvertex}
\Gamma^{\text{3g}}_{\mu\nu\rho}(p_1, p_2, p_3) = \mathcal{A}(s_0)\cdot\Gamma^{(0)}_{\mu\nu\rho}(p_1, p_2, p_3)\;.
\end{align}
The dressing function $\mathcal{A}(s_0)$ is obtained by solving the three-gluon 
vertex DSE in a `ghost triangle' approximation, depicted in 
Fig.~\ref{fig:3gvertexghostonly}. The details of the calculation can be found 
in~\cite{Eichmann:2014xya}. The resultant dressing function is shown 
in Fig.~\ref{fig:threegluon}. Information available from continuum 
non-perturbative studies of the three-gluon 
vertex~\cite{Binosi:2013rba,Aguilar:2013vaa,Blum:2014gna,Eichmann:2014xya} 
suggests that both the truncation of Fig.~\ref{fig:3gvertexghostonly}, and the 
restriction of possible tensor structures to the tree-level term, provide a 
reasonable phenomenological description of this Green's function. The effect 
which the dressed three-gluon vertex has on the hadron masses can be seen in Table~\ref{tab:comparewithlattice}. 

There is one further extension to our model that is possible, which is the inclusion 
of the so-called Abelian diagram in the quark-gluon vertex DSE, see Ref.~\cite{Williams:2014iea}. 
This introduces no complications in the evaluation of the quark-gluon vertex itself, and through the `cutting' 
procedure it is straightforward to construct a solvable BSE kernel which is consistent with axWTI \cite{Bender:1996bb}.
This BSE kernel would contain diagram with a new topology -- the so-called crossed ladder
diagram -- which increases the algebraic and numerical effort considerably. However, in previous calculations
the Abelian contribution has been shown to have a small effect on meson masses, typically less than 
two percent~\cite{Williams:2009wx}, which would similarly apply to our present investigation.
For these reasons, and in light of other uncertainties of continuum and lattice investigations, we feel that it 
is justified to ignore this extension for now. 

\subsection{Discussion}\label{sec:analysis}
Comparison of our results with the 
lattice~\cite{Lewis:2011zb, Hietanen:2014xca} requires the scale to be set by 
equating the electroweak (EW) scale with the pseudoscalar meson (`pion') decay 
constant, i.e. $v_{\mathrm{EW}} = f_{\pi} =246$~GeV. This puts the theory under 
investigation in the context of dynamical EW symmetry breaking, otherwise known 
as Technicolor (TC) \cite{Weinberg:1979bn, Dimopoulos:1979es}. 

The drawbacks that the SU(2) model discussed here (and any other model with QCD-like dynamics) faces as a Technicolor
template are by now well known. These include the problems with precision tests on
flavor-changing neutral currents~\cite{Eichten:1979ah}, and the composite `Higgs 
boson' which is expected to be very heavy. This latter problem is seen here, 
whereupon we do not find an isoscalar scalar (`sigma') meson
(a TC version of the Higgs boson) below $1.33$~TeV. This situation, however,
might change drastically if one considers explicitly the couplings to Standard Model 
particles \cite{Foadi:2012bb}, or more general EW embeddings~\cite{Cacciapaglia:2014uja}. 
Another promising approach to Technicolor phenomenology is to use nearly conformal theories
as Technicolor templates \cite{Holdom:1981rm,Holdom:1984sk,Akiba:1985rr,Yamawaki:1985zg}.
As we are presently concerned with the QCD-like aspects of the model under investigation,
we will not comment on its possible Technicolor applications further.      

Ground state masses for various $J^{PC}$ mesons are shown for both the
rainbow-ladder (RL) and beyond rainbow-ladder (BRL) truncations in 
Table~\ref{tab:comparewithlattice}, where they are additionally compared with 
the relevant lattice calculations. RL results were obtained by means of direct analytic 
continuation, whilst those of BRL were extrapolated from the region of spacelike 
$P^2$ via the inverse vertex function. The pion decay constant, which is used to set 
the scale of the calculation, is evaluated via the relation~\cite{Tandy:1997qf}:  
\begin{align}\label{eqn:piondecay}
f_\pi = \frac{Z_2 N_c}{\sqrt{2}P^2}\text{tr}\int_k\Gamma_\pi(k,-P)S(k_+)\gamma_5\slashed{P}S(k_-),
\end{align}
where $k_\pm = k \pm P/2$ and $\Gamma_\pi$ is the pion BSE amplitude normalised according
to the Nakanishi condition \cite{Nakanishi:1965zz}. In QCD, the conventions employed in the 
above equation would correspond to the value $f_\pi = 93$~MeV. When working in the 
region of spacelike $P^2$, the definition of Eq.~\eqref{eqn:piondecay} can be used without
approximations only in the chiral limit, since the pion amplitude $\Gamma_\pi$ can be obtained 
for the case $P^2 \rightarrow 0$. For non-chiral quarks, and thus non-vanishing pion mass, the 
calculation would have to be set up for complex total momentum, which is a formidable task in 
a BRL setting \cite{SanchisAlepuzWilliams:2014}. 

\begin{figure}[!t]
\begin{center}
\includegraphics[width=0.99\columnwidth]{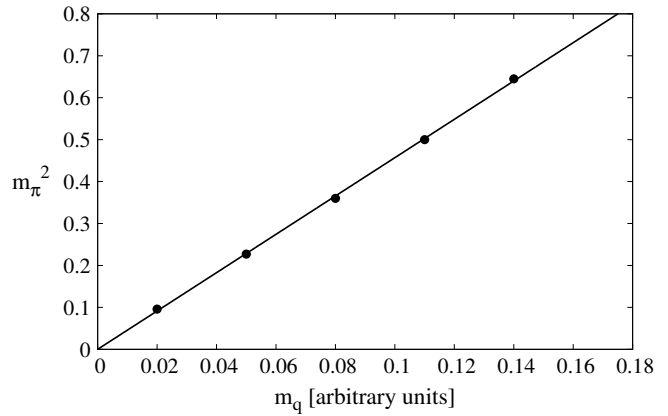}
\caption{Adherence of the calculated pion mass (squared) to the GMOR relation, as a function
of the (corrected) quark mass $m_q$. }\label{fig:gmor}
\end{center}
\end{figure}

\begin{table*}[!t]
\begin{center}
\caption{Chiral limit results for meson masses in rainbow-ladder (RL) and beyond rainbow-ladder
(BRL) truncations, compared with lattice data for an SU(2) theory. All units are in
TeV. Errors of the BRL results come from the extrapolation procedure. For the $0^{++}$ state, 
our continuum result is for an isoscalar; lattice results are forthcoming.}
\label{tab:comparewithlattice}
\begin{tabular}{c|c|c|c|c}
\hline\hline\noalign{\smallskip} 
$J^{PC}$    &    RL      & BRL, bare 3g vertex & BRL, dressed 3g vertex & Lattice, from \cite{Lewis:2011zb, Hietanen:2014xca} \\
\noalign{\smallskip}\hline
$0^{-+}$    &    0       &        $ 0          $           &        $0            $     &       --  \\      
$0^{++}$    &   1.24     &        $1.39(6)     $           &        $1.33(6)      $     &       --  \\
\noalign{\smallskip}\hline
$1^{--}$    &   1.95     &        $2.27(9)     $           &        $2.36(8)      $     &       $2.5 \pm 0.5$  \\
$1^{++}$    &   2.36     &        $2.87(10)    $           &        $3.08(10)     $     &       $3.3 \pm 0.7$  \\
\noalign{\smallskip}\hline\hline 
\end{tabular}
\end{center}
\end{table*}

For the discussion of results it would be useful to have an estimate on the mass of an 
isoscalar scalar meson, calculated in a method different from our DSE/BSE approach.
Since the lattice results for this particle are yet to come, we will use the values obtained 
by means of group theory scaling, which for the model under investigation gives $m_{\sigma}\in$
$\left[1,1.5\right]$~TeV~\cite{Foadi:2012bb}. Taking this into consideration, it seems that the RL method 
fares well for the sigma meson, and to a lesser extent, the rho meson. In the $1^{++}$
channel, this truncation performs inadequately, with a result which deviates by about 30 percent 
from the central lattice value. It is arguable whether or not one can modify the RL method so that it is
better suited for an SU(2) theory, thus performing reasonably well for all considered mesons. Based on our
current results, and given the limitations of the RL framework, we are skeptical towards this prospect.  

On the other hand, the results of the BRL approach compare well with lattice, especially when employing the dressed
three-gluon vertex. Since there are considerable error margins present in both the continuum and  lattice investigations,
stronger statements about the agreement of our methods will have to wait for more refined  calculations.
  
\begin{figure}[!b]
\begin{center}
\includegraphics[width=0.99\columnwidth]{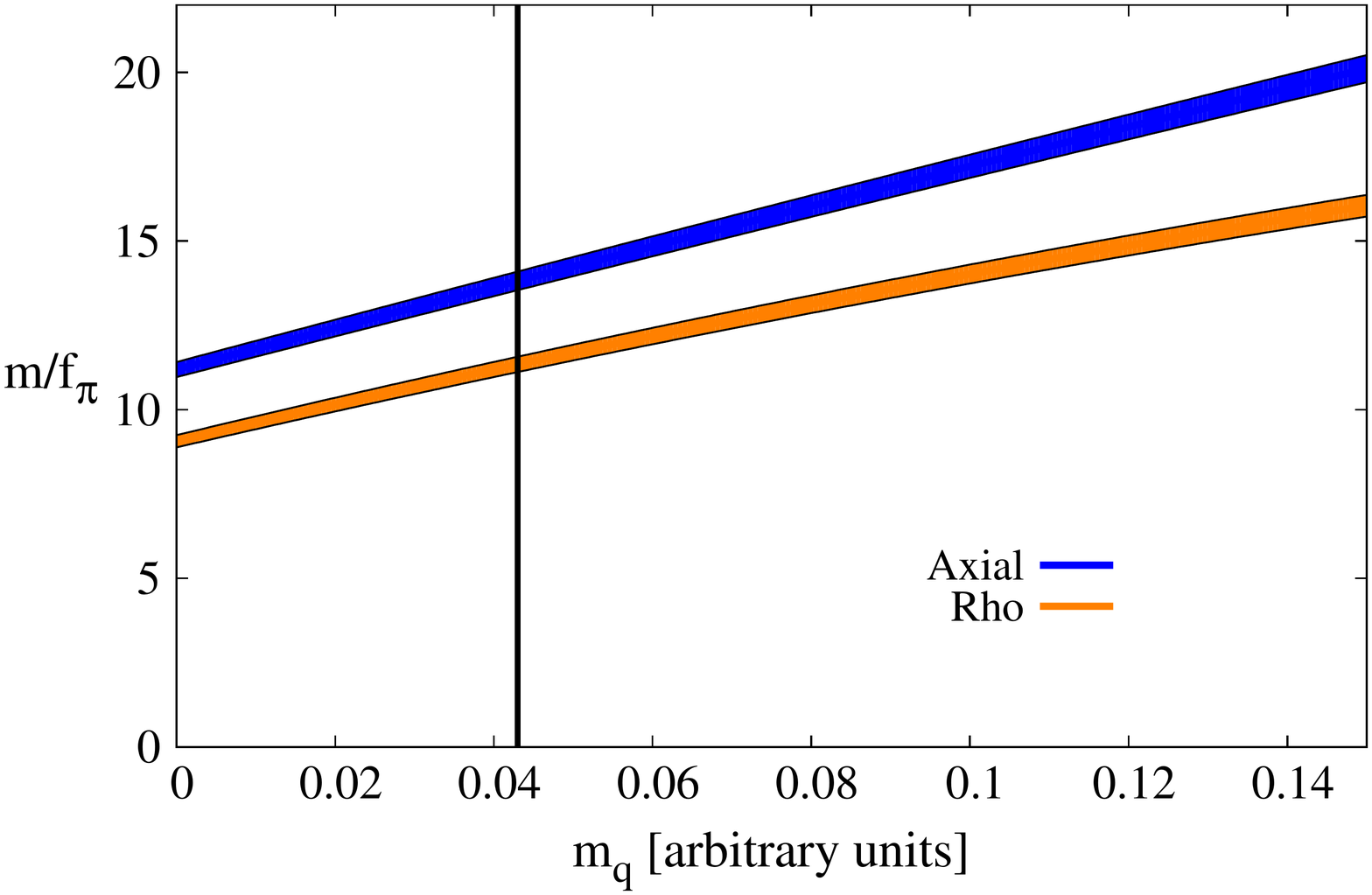}
\caption{$J=1$ meson masses (in units of chiral limit $f_\pi$) as a function of current quark 
mass. Bands correspond to uncertainties due to the extrapolation. The right-hand 
side of the vertical line corresponds to the region where $m_\rho \leq 2m_\pi$.}\label{fig:rhoaxial}
\end{center}
\end{figure}

Regarding the continuum calculation, dressing of the three-gluon vertex seems
to lead to a better agreement with the discretised approach, but the overall impact
of this modification is relatively mild, and all meson masses are rather 
robust in this respect. This leaves open the possibility that more elaborate modifications
of our model (i. e. inclusion of additional diagrams and higher $n$-point Green's functions 
in the quark-gluon vertex DSE) might not change the results appreciably. However, note that
dynamical contributions that can collectively be termed `pion cloud' effects are known to 
be important, and are the focus of present and future investigations.

Aside from the chiral limit study, we also performed calculations for 
non-vanishing current quark masses. In Fig.~\ref{fig:gmor} we demonstrate the 
validity of Gell-Mann-Oakes-Renner (GMOR) relation in the BRL approach,
while in Fig.~\ref{fig:rhoaxial} we plot the masses of spin one mesons (in units of
chiral limit $f_\pi$) as a function of current quark mass.
Both plots correspond to a calculation with a bare three-gluon vertex. The results shown in
Fig.~\ref{fig:rhoaxial} seem to compare well with the ones shown in Fig.~6 of \cite{Hietanen:2014xca}: 
however, a direct comparison is not possible since we don't have enough information
to relate our $m_q$ to the ones employed in \cite{Hietanen:2014xca}.  

As a final remark, we note that the calculation of the baryonic spectrum in this theory does not 
require any additional effort. An SU(2) gauge theory possesses an enlarged (Pauli-G{\"u}rsey)
flavor symmetry, which implies that chiral 
multiplets will contain both mesons and baryons (diquarks). In other words, a 
meson with $J^P$ quantum numbers will be degenerate with a $J^{-P}$ diquark. 
This degeneracy (which breaks down if the chemical potential is raised above 
some critical value $\mu_c$) has been confirmed in numerous lattice 
investigations~\cite{Lewis:2011zb,Hietanen:2014xca,Hands:2000ei,Kogut:2001na,Aloisio:2000rb}.

\section{Conclusions and outlook}\label{sec:conclusion}
We presented a Dyson-Schwinger/Bethe-Salpeter calculation of ground state hadron 
masses in a theory with two colors and two fundamentally charged Dirac fermions. 
We employed a novel beyond rainbow-ladder method and obtained good agreement 
with lattice results for spin one mesons: however, improved calculations will be 
needed to reduce uncertainties in both lattice and continuum approaches. 

For $J=0$ mesons, we demonstrated that chiral dynamics are satisfied (i.e. the 
GMOR relation holds) and obtained the mass of the sigma meson
in good agreement with the analysis based on group theory scaling. 
Additionally, we showed that the rainbow-ladder method performs unsatisfactorily
in this strongly-interacting template. This underlines the need to use more sophisticated
techniques when studying generic non-Abelian gauge theories.  

Besides masses, the beyond rainbow-ladder approach we outlined here can also be
used to study hadronic decays and form factors. A first step towards accessing these
quantities is to extend the calculation to complex Euclidean momenta. However, 
the technical complications which arise are considerable and are subject to
future investigation. 

\section*{Acknowledgments}
We would like to thank R.~Alkofer, C.~S.~Fischer, A.~Maas, H.~Sanchis-Alepuz, 
and F.~Sannino for useful discussions and a critical reading of this manuscript. 
This work was supported by the Helmholtz International Center for FAIR within 
the LOEWE program of the State of Hesse, a Lise-Meitner fellowship M1333--N16 
from the Austrian Science Fund (FWF), and from the Doktoratskolleg 
``Hadrons in Vacuum, Nuclei and Stars'' (FWF) DK W1203-N16. 
Further support by the  European Union (Hadron Physics 3 project ``Exciting Physics 
of Strong Interactions'') is acknowledged.


\begin{thebibliography}{99}
\bibitem{Hands:1999md}
  S.~Hands, J.~B.~Kogut, M.~P.~Lombardo and S.~E.~Morrison,
    Nucl.\ Phys.\ B {\bf 558} (1999) 327
  [hep-lat/9902034].
      

\bibitem{Aloisio:2000rb}
  R.~Aloisio, V.~Azcoiti, G.~Di Carlo, A.~Galante and A.~F.~Grillo,
    Nucl.\ Phys.\ B {\bf 606} (2001) 322
  [hep-lat/0011079].
    

\bibitem{Aloisio:2000if}
  R.~Aloisio, V.~Azcoiti, G.~Di Carlo, A.~Galante and A.~F.~Grillo,
    Phys.\ Lett.\ B {\bf 493} (2000) 189
  [hep-lat/0009034].
  

\bibitem{Hands:2000ei}   
  S.~Hands, I.~Montvay, S.~Morrison, M.~Oevers, L.~Scorzato and J.~Skullerud,
    Eur.\ Phys.\ J.\ C {\bf 17} (2000) 285
  [hep-lat/0006018].  
  

\bibitem{Kogut:2001na}
  J.~B.~Kogut, D.~K.~Sinclair, S.~J.~Hands and S.~E.~Morrison,
    Phys.\ Rev.\ D {\bf 64} (2001) 094505
  [hep-lat/0105026].
      

\bibitem{Hands:2001ee}
  S.~Hands, I.~Montvay, L.~Scorzato and J.~Skullerud,
    Eur.\ Phys.\ J.\ C {\bf 22} (2001) 451
  [hep-lat/0109029].
  

\bibitem{Kogut:2002cm}
  J.~B.~Kogut, D.~Toublan and D.~K.~Sinclair,
    Nucl.\ Phys.\ B {\bf 642} (2002) 181
  [hep-lat/0205019].    

\bibitem{Muroya:2002ry}
  S.~Muroya, A.~Nakamura and C.~Nonaka,
    Phys.\ Lett.\ B {\bf 551} (2003) 305
  [hep-lat/0211010].
    

\bibitem{Alles:2006ea}                          
  B.~Alles, M.~D'Elia and M.~P.~Lombardo,
    Nucl.\ Phys.\ B {\bf 752} (2006) 124
  [hep-lat/0602022].

\bibitem{Hands:2006ve}
  S.~Hands, S.~Kim and J.~I.~Skullerud,
    Eur.\ Phys.\ J.\ C {\bf 48} (2006) 193
  [hep-lat/0604004].
    

\bibitem{Cotter:2012mb}
  S.~Cotter, P.~Giudice, S.~Hands and J.~I.~Skullerud,
    Phys.\ Rev.\ D {\bf 87} (2013) 3,  034507
  [arXiv:1210.4496 [hep-lat]].
    

\bibitem{Lewis:2011zb}
  R.~Lewis, C.~Pica and F.~Sannino,
    Phys.\ Rev.\ D {\bf 85} (2012) 014504
  [arXiv:1109.3513 [hep-ph]].
    

\bibitem{Hietanen:2014xca}
  A.~Hietanen, R.~Lewis, C.~Pica and F.~Sannino,
    JHEP {\bf 1407} (2014) 116
  [arXiv:1404.2794 [hep-lat]].
    

\bibitem{Hietanen:2013fya}
  A.~Hietanen, R.~Lewis, C.~Pica and F.~Sannino,
    arXiv:1308.4130 [hep-ph].
    

\bibitem{Roberts:1994dr}
  C.~D.~Roberts and A.~G.~Williams,
    Prog.\ Part.\ Nucl.\ Phys.\  {\bf 33} (1994) 477
  [hep-ph/9403224].
    

\bibitem{Alkofer:2000wg}
  R.~Alkofer and L.~von Smekal,
    Phys.\ Rept.\  {\bf 353} (2001) 281
  [hep-ph/0007355].
    

\bibitem{Fischer:2006ub}
  C.~S.~Fischer,
    J.\ Phys.\ G {\bf 32} (2006) R253
  [hep-ph/0605173].
    

\bibitem{Cloet:2013jya}
  I.~C.~Cloet and C.~D.~Roberts,
    Prog.\ Part.\ Nucl.\ Phys.\  {\bf 77} (2014) 1
  [arXiv:1310.2651 [nucl-th]].
    

\bibitem{Maris:1997hd}
  P.~Maris, C.~D.~Roberts and P.~C.~Tandy,
    Phys.\ Lett.\ B {\bf 420} (1998) 267
  [nucl-th/9707003].
    

\bibitem{Maris:1997tm}
  P.~Maris and C.~D.~Roberts,
    Phys.\ Rev.\ C {\bf 56} (1997) 3369
  [nucl-th/9708029].
    

\bibitem{Maris:1999nt}
  P.~Maris and P.~C.~Tandy,
    Phys.\ Rev.\ C {\bf 60} (1999) 055214
  [nucl-th/9905056].
    
    
\bibitem{Krassnigg:2009gd}
  A.~Krassnigg,
    PoS CONFINEMENT {\bf 8} (2008) 075
  [arXiv:0812.3073 [nucl-th]].
    

\bibitem{Krassnigg:2009zh}
  A.~Krassnigg,
    Phys.\ Rev.\ D {\bf 80} (2009) 114010
  [arXiv:0909.4016 [hep-ph]].
    

\bibitem{Blank:2010pa}
  M.~Blank, A.~Krassnigg and A.~Maas,
    Phys.\ Rev.\ D {\bf 83} (2011) 034020
  [arXiv:1007.3901 [hep-ph]].
    

\bibitem{Krassnigg:2010mh}
  A.~Krassnigg and M.~Blank,
    Phys.\ Rev.\ D {\bf 83} (2011) 096006
  [arXiv:1011.6650 [hep-ph]].
    

\bibitem{Blank:2011ha}
  M.~Blank and A.~Krassnigg,
    Phys.\ Rev.\ D {\bf 84} (2011) 096014
  [arXiv:1109.6509 [hep-ph]].
    

\bibitem{Roberts:2011cf}
  H.~L.~L.~Roberts, L.~Chang, I.~C.~Cloet and C.~D.~Roberts,
    Few Body Syst.\  {\bf 51} (2011) 1
  [arXiv:1101.4244 [nucl-th]].
    

\bibitem{Fischer:2014xha}
  C.~S.~Fischer, S.~Kubrak and R.~Williams,
    Eur.\ Phys.\ J.\ A {\bf 50} (2014) 126
  [arXiv:1406.4370 [hep-ph]].
    

\bibitem{Hilger:2014nma}
  T.~Hilger, C.~Popovici, M.~Gomez-Rocha and A.~Krassnigg,
    arXiv:1409.3205 [hep-ph].
    

\bibitem{Fischer:2014cfa}
  C.~S.~Fischer, S.~Kubrak and R.~Williams,
    arXiv:1409.5076 [hep-ph].
    

\bibitem{Eichmann:2009qa}
  G.~Eichmann, R.~Alkofer, A.~Krassnigg and D.~Nicmorus,
    Phys.\ Rev.\ Lett.\  {\bf 104} (2010) 201601
  [arXiv:0912.2246 [hep-ph]].
    

\bibitem{Nicmorus:2010sd}
  D.~Nicmorus, G.~Eichmann and R.~Alkofer,
    Phys.\ Rev.\ D {\bf 82} (2010) 114017
  [arXiv:1008.3184 [hep-ph]].
    

\bibitem{SanchisAlepuz:2011aa}
  H.~Sanchis-Alepuz, R.~Alkofer, G.~Eichmann and R.~Williams,
    PoS QCD {\bf -TNT-II} (2011) 041
  [arXiv:1112.3214 [hep-ph]].
    

\bibitem{Eichmann:2011vu}
  G.~Eichmann,
    Phys.\ Rev.\ D {\bf 84} (2011) 014014
  [arXiv:1104.4505 [hep-ph]].
    

\bibitem{Sanchis-Alepuz:2013iia}
  H.~Sanchis-Alepuz, R.~Williams and R.~Alkofer,
    Phys.\ Rev.\ D {\bf 87} (2013) 9,  096015
  [arXiv:1302.6048 [hep-ph]].
    

\bibitem{Munczek:1994zz}
  H.~J.~Munczek,
    Phys.\ Rev.\ D {\bf 52} (1995) 4736
  [hep-th/9411239].
    

\bibitem{Bender:1996bb}
  A.~Bender, C.~D.~Roberts and L.~Von Smekal,
    Phys.\ Lett.\ B {\bf 380} (1996) 7
  [nucl-th/9602012].
    

\bibitem{Bender:2002as}
  A.~Bender, W.~Detmold, C.~D.~Roberts and A.~W.~Thomas,
    Phys.\ Rev.\ C {\bf 65} (2002) 065203
  [nucl-th/0202082].
    

\bibitem{Watson:2004kd}
  P.~Watson, W.~Cassing and P.~C.~Tandy,
    Few Body Syst.\  {\bf 35} (2004) 129
  [hep-ph/0406340].
    

\bibitem{Watson:2004jq}
  P.~Watson and W.~Cassing,
    Few Body Syst.\  {\bf 35} (2004) 99
  [hep-ph/0405287].
    

\bibitem{Bhagwat:2004hn}
  M.~S.~Bhagwat, A.~Holl, A.~Krassnigg, C.~D.~Roberts and P.~C.~Tandy,
    Phys.\ Rev.\ C {\bf 70} (2004) 035205
  [nucl-th/0403012].
    

\bibitem{Matevosyan:2006bk}
  H.~H.~Matevosyan, A.~W.~Thomas and P.~C.~Tandy,
    Phys.\ Rev.\ C {\bf 75} (2007) 045201
  [nucl-th/0605057].
   
   
\bibitem{Fischer:2009jm}
  C.~S.~Fischer and R.~Williams,
    Phys.\ Rev.\ Lett.\  {\bf 103} (2009) 122001
  [arXiv:0905.2291 [hep-ph]].
    

\bibitem{Windisch:2012de}
  A.~Windisch, M.~Hopfer and R.~Alkofer,
    Acta Phys.\ Polon.\ Supp.\  {\bf 6} (2013) 347
  [arXiv:1210.8428 [hep-ph]].
    

\bibitem{Gomez-Rocha:2014vsa}
  M.~Gomez-Rocha, T.~Hilger and A.~Krassnigg,
    arXiv:1408.1077 [hep-ph].
    

\bibitem{Williams:2014iea}
  R.~Williams,
    arXiv:1404.2545 [hep-ph].
    

\bibitem{Fischer:2005en}
  C.~S.~Fischer, P.~Watson and W.~Cassing,
    Phys.\ Rev.\ D {\bf 72} (2005) 094025
  [hep-ph/0509213].
    

\bibitem{Fischer:2007ze}
  C.~S.~Fischer, D.~Nickel and J.~Wambach,
    Phys.\ Rev.\ D {\bf 76} (2007) 094009
  [arXiv:0705.4407 [hep-ph]].
    

\bibitem{Fischer:2008sp}
  C.~S.~Fischer, D.~Nickel and R.~Williams,
    Eur.\ Phys.\ J.\ C {\bf 60} (2009) 47
  [arXiv:0807.3486 [hep-ph]].
    

\bibitem{Fischer:2008wy}
  C.~S.~Fischer and R.~Williams,
    Phys.\ Rev.\ D {\bf 78} (2008) 074006
  [arXiv:0808.3372 [hep-ph]].
    

\bibitem{Chang:2009zb}
  L.~Chang and C.~D.~Roberts,
    Phys.\ Rev.\ Lett.\  {\bf 103} (2009) 081601
  [arXiv:0903.5461 [nucl-th]].
    

\bibitem{Chang:2011ei}
  L.~Chang and C.~D.~Roberts,
    Phys.\ Rev.\ C {\bf 85} (2012) 052201
  [arXiv:1104.4821 [nucl-th]].
    

\bibitem{Heupel:2014ina}
  W.~Heupel, T.~Goecke and C.~S.~Fischer,
    Eur.\ Phys.\ J.\ A {\bf 50} (2014) 85
  [arXiv:1402.5042 [hep-ph]].
    

\bibitem{vonSmekal:2009ae}
  L.~von Smekal, K.~Maltman and A.~Sternbeck,
    Phys.\ Lett.\ B {\bf 681} (2009) 336
  [arXiv:0903.1696 [hep-ph]].
    

\bibitem{Eichmann:2008ae}
  G.~Eichmann, R.~Alkofer, I.~C.~Cloet, A.~Krassnigg and C.~D.~Roberts,
    Phys.\ Rev.\ C {\bf 77} (2008) 042202
  [arXiv:0802.1948 [nucl-th]].
    

\bibitem{Qin:2011xq}
  S.~x.~Qin, L.~Chang, Y.~x.~Liu, C.~D.~Roberts and D.~J.~Wilson,
    Phys.\ Rev.\ C {\bf 85} (2012) 035202
  [arXiv:1109.3459 [nucl-th]].
    

\bibitem{Skullerud:2002ge}
  J.~Skullerud and A.~Kizilersu,
    JHEP {\bf 0209} (2002) 013
  [hep-ph/0205318].
    

\bibitem{Skullerud:2003qu}
  J.~I.~Skullerud, P.~O.~Bowman, A.~Kizilersu, D.~B.~Leinweber and A.~G.~Williams,
    JHEP {\bf 0304} (2003) 047
  [hep-ph/0303176].
    

\bibitem{Kizilersu:2006et}
  A.~Kizilersu, D.~B.~Leinweber, J.~I.~Skullerud and A.~G.~Williams,
    Eur.\ Phys.\ J.\ C {\bf 50} (2007) 871
  [hep-lat/0610078].
    

\bibitem{Rojas:2013tza}
  E.~Rojas, J.~P.~B.~C.~de Melo, B.~El-Bennich, O.~Oliveira and T.~Frederico,
    JHEP {\bf 1310} (2013) 193
  [arXiv:1306.3022 [hep-ph]].
    

\bibitem{Aguilar:2014lha}
  A.~C.~Aguilar, D.~Binosi, D.~Iba\~nez and J.~Papavassiliou,
    Phys.\ Rev.\ D {\bf 90} (2014) 065027
  [arXiv:1405.3506 [hep-ph]].
      

\bibitem{Mitter:2014wpa}
  M.~Mitter, J.~M.~Pawlowski and N.~Strodthoff,
  Phys.\ Rev.\ D {\bf 91} (2015) 054035
  [arXiv:1411.7978 [hep-ph]].



\bibitem{Eichmann:2014xya}
  G.~Eichmann, R.~Williams, R.~Alkofer and M.~Vujinovic,
    Phys.\ Rev.\ D {\bf 89} (2014) 105014
  [arXiv:1402.1365 [hep-ph]].
    

\bibitem{Fischer:2003rp}
  C.~S.~Fischer and R.~Alkofer,
    Phys.\ Rev.\ D {\bf 67} (2003) 094020
  [hep-ph/0301094].
    

\bibitem{Caswell:1974gg}         
  W.~E.~Caswell,
    Phys.\ Rev.\ Lett.\  {\bf 33} (1974) 244.  
  
 

\bibitem{Appelquist:1986an}
  T.~W.~Appelquist, D.~Karabali and L.~C.~R.~Wijewardhana,
    Phys.\ Rev.\ Lett.\  {\bf 57} (1986) 957.
  

\bibitem{Appelquist:1987fc}
  T.~Appelquist and L.~C.~R.~Wijewardhana,
    Phys.\ Rev.\ D {\bf 36} (1987) 568.
  
 

\bibitem{Appelquist:2007hu}
  T.~Appelquist, G.~T.~Fleming and E.~T.~Neil,
    Phys.\ Rev.\ Lett.\  {\bf 100} (2008) 171607
   [Erratum-ibid.\  {\bf 102} (2009) 149902]
  [arXiv:0712.0609 [hep-ph]].  
  

\bibitem{Bursa:2010xn}
  F.~Bursa, L.~Del Debbio, L.~Keegan, C.~Pica and T.~Pickup,
    Phys.\ Lett.\ B {\bf 696} (2011) 374
  [arXiv:1007.3067 [hep-ph]].  
    
\bibitem{Hasenfratz:2011xn}
  A.~Hasenfratz,
    Phys.\ Rev.\ Lett.\  {\bf 108} (2012) 061601
  [arXiv:1106.5293 [hep-lat]].  
  

\bibitem{Cheng:2011ic}                              
  A.~Cheng, A.~Hasenfratz and D.~Schaich,
    Phys.\ Rev.\ D {\bf 85} (2012) 094509
  [arXiv:1111.2317 [hep-lat]].  
  

\bibitem{Lin:2012iw}
  C.-J.~D.~Lin, K.~Ogawa, H.~Ohki and E.~Shintani,
    JHEP {\bf 1208} (2012) 096
  [arXiv:1205.6076 [hep-lat]].
  

\bibitem{Hopfer:2014zna}
  M.~Hopfer, C.~S.~Fischer and R.~Alkofer,
    JHEP {\bf 1411} (2014) 035
  [arXiv:1405.7031 [hep-ph]]. 
  

\bibitem{Sannino:2004qp}             
  F.~Sannino and K.~Tuominen,
    Phys.\ Rev.\ D {\bf 71} (2005) 051901
  [hep-ph/0405209].
  

\bibitem{Dietrich:2005jn}
  D.~D.~Dietrich, F.~Sannino and K.~Tuominen,
    Phys.\ Rev.\ D {\bf 72} (2005) 055001
  [hep-ph/0505059].

\bibitem{Dietrich:2006cm}
  D.~D.~Dietrich and F.~Sannino,
  Phys.\ Rev.\ D {\bf 75} (2007) 085018
  [hep-ph/0611341].
  
\bibitem{Catterall:2007yx}
  S.~Catterall and F.~Sannino,
  Phys.\ Rev.\ D {\bf 76} (2007) 034504
  [arXiv:0705.1664 [hep-lat]].

\bibitem{Maas:2011jf}
  A.~Maas,
  JHEP {\bf 1105} (2011) 077
  [arXiv:1102.5023 [hep-lat]].

\bibitem{DeGrand:2011qd}
  T.~DeGrand, Y.~Shamir and B.~Svetitsky,
    Phys.\ Rev.\ D {\bf 83} (2011) 074507
  [arXiv:1102.2843 [hep-lat]].    
  

\bibitem{Bursa:2011ru}
  F.~Bursa, L.~Del Debbio, D.~Henty, E.~Kerrane, B.~Lucini, A.~Patella, C.~Pica and T.~Pickup {\it et al.},
    Phys.\ Rev.\ D {\bf 84} (2011) 034506
  [arXiv:1104.4301 [hep-lat]].   
  

\bibitem{Fodor:2012ty}
  Z.~Fodor, K.~Holland, J.~Kuti, D.~Nogradi, C.~Schroeder and C.~H.~Wong,
    Phys.\ Lett.\ B {\bf 718} (2012) 657
  [arXiv:1209.0391 [hep-lat]].   
   

\bibitem{DeGrand:2013uha}
  T.~DeGrand, Y.~Shamir and B.~Svetitsky,
    Phys.\ Rev.\ D {\bf 88} (2013) 5,  054505
  [arXiv:1307.2425].

  
\bibitem{GimenoSegovia:2008sx}
  M.~Gimeno-Segovia and F.~J.~Llanes-Estrada,
    Eur.\ Phys.\ J.\ C {\bf 56} (2008) 557
  [arXiv:0805.4145 [hep-th]].
    

\bibitem{Strauss:2012dg}
  S.~Strauss, C.~S.~Fischer and C.~Kellermann,
    Phys.\ Rev.\ Lett.\  {\bf 109} (2012) 252001
  [arXiv:1208.6239 [hep-ph]].
    

\bibitem{Windisch:2013dxa}
  A.~Windisch, M.~Q.~Huber and R.~Alkofer,
    Acta Phys.\ Polon.\ Supp.\  {\bf 6} (2013) 3,  887
  [arXiv:1304.3642 [hep-ph]].
    

\bibitem{Bhagwat:2007rj}
  M.~S.~Bhagwat, A.~Hoell, A.~Krassnigg, C.~D.~Roberts and S.~V.~Wright,
    Few Body Syst.\  {\bf 40} (2007) 209
  [nucl-th/0701009].
    

\bibitem{Dorkin:2013rsa}
  S.~M.~Dorkin, L.~P.~Kaptari, T.~Hilger and B.~Kampfer,
    Phys.\ Rev.\ C {\bf 89} (2014) 3,  034005
  [arXiv:1312.2721 [hep-ph]].
    

\bibitem{Berges:2004pu}
  J.~Berges,
    Phys.\ Rev.\ D {\bf 70} (2004) 105010
  [hep-ph/0401172].
    

\bibitem{Binosi:2013rba}
  D.~Binosi, D.~Iba\~nez and J.~Papavassiliou,
    Phys.\ Rev.\ D {\bf 87} (2013) 12,  125026
  [arXiv:1304.2594 [hep-ph]].
    

\bibitem{Aguilar:2013vaa}
  A.~C.~Aguilar, D.~Binosi, D.~Iba\~nez and J.~Papavassiliou,
    Phys.\ Rev.\ D {\bf 89} (2014) 085008
  [arXiv:1312.1212 [hep-ph]].
    

\bibitem{Blum:2014gna}
  A.~Blum, M.~Q.~Huber, M.~Mitter and L.~von Smekal,
    Phys.\ Rev.\ D {\bf 89} (2014) 061703
  [arXiv:1401.0713 [hep-ph]].

\bibitem{Williams:2009wx}
  R.~Williams,
  EPJ Web Conf.\  {\bf 3} (2010) 03005
  [arXiv:0912.3494 [hep-ph]].
 

\bibitem{Weinberg:1979bn}
  S.~Weinberg,
    Phys.\ Rev.\ D {\bf 19} (1979) 1277.
    

\bibitem{Dimopoulos:1979es}
  S.~Dimopoulos and L.~Susskind,
    Nucl.\ Phys.\ B {\bf 155} (1979) 237.
    

\bibitem{Eichten:1979ah}
  E.~Eichten and K.~D.~Lane,
    Phys.\ Lett.\ B {\bf 90} (1980) 125.
    

\bibitem{Foadi:2012bb}
  R.~Foadi, M.~T.~Frandsen and F.~Sannino,
  Phys.\ Rev.\ D {\bf 87} (2013) 9,  095001
  [arXiv:1211.1083 [hep-ph]].

\bibitem{Cacciapaglia:2014uja}
  G.~Cacciapaglia and F.~Sannino,
  JHEP {\bf 1404} (2014) 111
  [arXiv:1402.0233 [hep-ph]].

\bibitem{Holdom:1981rm}                                          
  B.~Holdom,
    Phys.\ Rev.\ D {\bf 24} (1981) 1441.
      

\bibitem{Holdom:1984sk}
  B.~Holdom,
    Phys.\ Lett.\ B {\bf 150} (1985) 301.
    

\bibitem{Akiba:1985rr}
  T.~Akiba and T.~Yanagida,
    Phys.\ Lett.\ B {\bf 169} (1986) 432.
  

\bibitem{Yamawaki:1985zg}
  K.~Yamawaki, M.~Bando and K.~i.~Matumoto,
    Phys.\ Rev.\ Lett.\  {\bf 56} (1986) 1335.
    
  
\bibitem{Tandy:1997qf}
  P.~C.~Tandy,
  Prog.\ Part.\ Nucl.\ Phys.\  {\bf 39} (1997) 117
  [nucl-th/9705018]. 
  
  
\bibitem{Nakanishi:1965zz}
  N.~Nakanishi,
  Phys.\ Rev.\  {\bf 139} (1965) B1401.

  
\bibitem{SanchisAlepuzWilliams:2014}
  H.~Sanchis-Alepuz and R.~Williams
  \emph{in preparation}.
    
\end{thebibliography}
\end{document}